\documentstyle [aps]{revtex}

\newcommand {\beq}{\equation}
\newcommand {\eeq}{\endequation}
\newcommand {\beqa}{\begin{eqnarray}}
\newcommand {\eeqa}{\end{eqnarray}}
\newcommand {\la} {\langle}
\newcommand {\ra} {\rangle}
\newcommand {\rl} {\rangle\langle}
\newcommand {\inid} {\int\limits_0^\infty d}

\begin {document}

\title {Oscillating decay of an unstable system}
\author {I.~Antoniou$^1$, E.~Karpov$^1$, G.~Pronko$^{1,2}$, and 
E.~Yarevsky$^{1,3}$}
\address{$^1$ International Solvay Institutes for Physics and Chemistry,
Campus Plaine ULB, C.P. 231, Bd. du Triomphe, Brussels B-1050, Belgium}
\address{$^2$ Institute for High Energy Physics, Protvino, Moscow region 
142284, Russia}
\address{$^3$ Laboratory of Complex Systems Theory, Institute for Physics,
St.Petersburg State University, Uljanovskaya~1, St.Petersburg 198904, Russia}
\date {\today}

\maketitle

\begin {abstract}
We study the short-time and medium-time behavior of the survival 
probability in the frame of the $N$-level Friedrichs model. 
The time evolution of an arbitrary unstable initial state is determined.
We show that the survival probability may oscillate significantly
during the so-called exponential era. This result explains qualitatively 
the experimental observations of the NaI decay.
\end {abstract}

\pacs{03.65.Bz}

\section {Introduction}

Recent developments in femtosecond laser optics,
see for example the XXth Solvay Conference on Chemistry~\cite {Solvay},
opened new possibilities for the study of quantum transitions, 
which are a very important subject of the quantum theory.
In a series of works, Zewail et al~\cite {Zewail1,Zewail2,Zewail3,Zewail4}
applied femtosecond transition-time spectroscopy
for the probing of chemical reactions. 
Following the work of Kinsey et al.~\cite {Kinsey},
they attempted in paper~\cite {Zewail4} to track wave packet trajectories
in the dissociation of NaI.

The shapes of the ground state potential for NaI and of the quasi-bound 
potential of the Na($^2$S$_{1/2}$)+I($^2$P$_{3/2}$) system
suggest a mechanism of the induced dissociation process.
The femtosecond laser pulse brings the NaI molecule to 
the state of quasi-bound ions. 
The distance between the ions reaches the region where two potentials 
have similar values due to vibrations of a NaI excited state. Then 
the transition from Na($^2$S$_{1/2}$)+I($^2$P$_{3/2}$) quasi-bound state
to NaI continuum state occurs resulting in the dissociation of the molecule.

After an initial exciting laser pulse, the experiment shows oscillations 
of the Na($^2$S$_{1/2}$)+I($^2$P$_{3/2}$) population, 
which are explained in~\cite {Zewail4} by wave packet propagation. 
The direction of the wave packet propagation
is correlated with the oscillation (extension  and contraction)
of the NaI bond. The quantum dynamics calculations are based on a
time-dependent perturbation formalism.

This problem is an example of the interaction of the discrete spectrum 
with the continuous spectrum, which was extensively discussed in the 
literature starting from the work of Friedrichs~\cite {Fried}. 
Indeed, the energy states of Na($^2$S$_{1/2}$)+I($^2$P$_{3/2}$) are the 
excited state embedded into the continuum states of the decay products.
Therefore, the time dependence of the Na($^2$S$_{1/2}$)+I($^2$P$_{3/2}$)
population is described by the survival probability of the excited state
prepared by the laser pulse.

The original Friedrichs model~\cite {Fried} contains two discrete 
energy level, a ground state and an excited state, coupled with the
continuum, being bounded from below. The time dependence of the survival 
probability of the excited state has been studied both 
theoretically~\cite {NPN,P2,AKPY,P3,kofman1,kofman2} and 
experimentally~\cite {Itano,Toschek,FGMR}.
It is exponential with a short non-exponential
initial era and a non-exponential long tail. 
As a result, Friedrichs models are very appropriate for the discussion 
of the particle decay and for the description of dressed unstable 
states~\cite {PPT,AP,OPPDUS}.
The analytical structure of the $N$-level Friedrichs model has been 
widely discussed~\cite {Fn1,Fn2,Fn3,Fn4,Fn5,Exner},
and the possibility of the oscillations of the survival
probability was pointed out in~\cite {kofman1,Fn1}.

In the present paper we shall show that the $N$-level Friedrichs model
can also explain the oscillations of the survival probability of the 
excited state observed by Zewail and co-workers~\cite {Zewail4}.
Several excited levels are necessary in order to construct a
wave packet, which can exhibit localization and
nonconventional time evolution. 
In Section~2 we present the model and describe the exact solution 
diagonalising the Hamiltonian. Using the relation between eigenstates
of the unperturbed Hamiltonian and the total Hamiltonian,
we describe in Section~3 the time evolution of the basis states.
Specifying the formfactor of the interaction, we show in Section~4 the
appearance of oscillations already for the two level Friedrichs model.
In Section~5 we demonstrate that the survival probability of unstable 
states in the $N$-level Friedrichs model is in fact very close to 
the one obtained in the experiment~\cite {Zewail4}.

\section {Model and exact solution}

The Hamiltonian of the Friedrichs model~\cite{Fried}
generalized to $N$-level is:
\beq
  H =  H_0 + \lambda V, 
\label{H}
\eeq
where
\begin{eqnarray}
\label {Ham2}
  H_0 & = & \sum_{k=1}^{N}\omega_k|k\rl k|
        + \int\limits_0^\infty d\omega\,\omega |\omega\rl \omega| ,
        \nonumber \\
V\,\, & = & \sum_{k=1}^N \int\limits_0^\infty d\omega f_k(\omega)
          \left(|k\rl \omega| + |\omega\rl k|\right).
\end{eqnarray}
Here $|k\ra$ represent states of the discrete spectrum with the energy 
$\omega_k$, $\omega_k >0$. We assume the simplest case that 
$\omega_k \neq \omega_{k'}$ for $k \ne k'$.
The vectors $|\omega\ra$ represent states of the continuous 
spectrum with the energy $\omega$, $f_k(\omega)$ are the formfactors for
the transitions between the discrete and the continuous spectrum, and
$\lambda$ is the coupling parameter. The vacuum energy is chosen to be zero. 
The states $|k\ra$ and $|\omega\ra$ form a complete orthonormal basis:
\equation \label{Ort}
  \la k| k'\ra   = \delta_{kk'}, \quad   
  \la\omega|\omega'\ra = \delta(\omega - \omega'), \quad  
  \la\omega|k\ra  = 0, \quad k,k'=1 \ldots N,
\endequation
\beq
\label {Comp}
  \sum_{k=1}^{N}\omega_k|k\rl k|
  + \int\limits_0^\infty d\omega\,|\omega\rl \omega| = I,
\eeq
where $\delta_{kk'}$ is the Kronecker symbol, 
$\delta(\omega - \omega')$ is the Dirac's delta function and
$I$ is the unity operator.
The Hamiltonian $H_0$ has the continuous spectrum on the 
interval $[0,\infty)$ and the discrete spectrum 
$\omega_1, ..., \omega_k$ embedded in the continuous spectrum. 

As the interaction $\lambda V$ is switched on, the eigenstates $|k\ra$ 
become resonances of $H$ as in the case of the one-level Friedrichs 
model~\cite {Fried}. Let us consider the eigenvalue problem for the 
$N$-level Friedrichs Hamiltonian (\ref{H})  
\equation \label{evp}
  H|\Psi_\omega \ra  =  \omega|\Psi_\omega \ra .
\endequation
We shall look for the solution of Eq.(\ref{evp}) in the form:
\beq \label{psitemp}
  |\Psi_\omega \ra 
  = \sum_k \psi_k(\omega)|k\ra + \inid\omega' \, \psi(\omega,\omega')|\omega'\ra,
\eeq
where $\psi_k(\omega)$ and $\psi(\omega,\omega')$ are unknown functions. 
Inserting (\ref{psitemp}) into (\ref{evp}) and making use of the 
orthogonality relations, we obtain for them a system of equations:
\beq \label {systeq}
  \left \{
  \begin{array}{l l l} 
        (\omega_k - \omega) \psi_k(\omega) 
        + \lambda \inid \omega' f_k(\omega') \psi(\omega,\omega') 
           & = & 0 \, , \\
       (\omega'- \omega)  \psi(\omega,\omega') 
        + { \displaystyle \lambda\sum_{k=1}^N } f_k(\omega')\psi_k(\omega)
            & = & 0 \, .
  \end{array}
  \right.
\eeq
Eliminating $\psi(\omega,\omega')$ from this system, we arrive at the 
following equation for $\psi_k(\omega)$:
\beq \label{Ao}
  \sum_{k'=1}^N G^{-1}_{k k '}(\omega) \psi_{k'}(\omega)
     = - C \lambda f_k(\omega),
\eeq
where $C$ is an arbitrary constant. $G_{kk'}(\omega)$ are the matrix elements
of the partial resolvent which is:
\beq \label {eta}
  G^{-1}_{k k '}(\omega)
    = (\omega_k - \omega)\delta_{k k'} - \lambda^2 \inid \omega'
      \frac{f_k(\omega')f_{k'}(\omega')}{\omega' - \omega}.
\eeq
Under certain conditions (which will be specified below, see 
also \cite {Exner}), the function 
$ G_{kk'}(z)$ is analytic everywhere in the first sheet of the Riemann 
manifold except for the cut $[0,\infty)$. In this case, the Hamiltonian 
$H$ has no discrete spectrum. The solution of Eq.(\ref{Ao}) is given by
\beq \label{Ao1}
 \psi_k(\omega)
     = - C \lambda \sum_{k'=1}^N G_{k k '} (\omega \pm i0) f_{k'}(\omega).
\eeq
With this equation we find $\psi(\omega,\omega')$ from the system (\ref{systeq}):
\beq \label{Aoo1}
  \psi(\omega,\omega')
  = C\Bigg[\delta(\omega - \omega') 
  +  \frac{\lambda^2 \sum_{k,k'=1}^N f_k(\omega')G_{k k'}(\omega \pm i0)
              f_{k'}(\omega)}{\omega - \omega' \pm i0} \Bigg] \, . 
\eeq
The eigenvalue problem (\ref{evp}) has two sets of solutions
\beq\label{evpsol}
  |\Psi_\omega \ra_{\stackrel{\rm in}{\rm out}}
    =  |\omega\ra + \lambda\sum_{k,l=1}^N f_l(\omega)G_{kl}(\omega \pm i0) \\
        \left\{\int\limits_0^{\infty}{\rm d}\omega'
                \frac{\lambda f_k(\omega')}
                     {\omega'-\omega \mp i0}|\omega'\ra - |k\ra
               \right\} ,
\eeq
which correspond to the ``in'' and ``out'' asymptotic conditions.
The value $C=1$ corresponds to the orthonormalization condition
\equation \label{orth}
  _{\stackrel{\rm in}{\rm out}}\la \Psi_\omega|
  \Psi_{\omega'}\ra_{\stackrel{\rm in}{\rm out}}
    = \delta(\omega-\omega') .
\endequation
We can also prove the completeness condition 
\equation \label{norm}
  \int\limits_0^\infty {\rm d}\omega\,
  |\Psi_\omega \ra_{\stackrel{\rm in}{\rm out}}
  {_{\stackrel{\rm in}{\rm out}}\la \Psi_\omega |}  = 
  \sum_{k=1}^{N}\omega_k|k\rl k|
  + \int\limits_0^\infty d\omega\,|\omega\rl \omega| .  
\endequation
Hence the new states diagonalize the total Hamiltonian (\ref{H}) as
\equation \label{Hdiagonal}
  H = \int\limits^{\infty}_0 {\rm d}\omega\, \omega
      |\Psi_\omega \ra_{\stackrel{\rm in}{\rm out}}
       {_{\stackrel{\rm in}{\rm out}}\la \Psi_\omega |} \, .
\endequation

The proof of completeness is based on the matrix formula
$$
  A^{-1} - B^{-1} = A^{-1}(B - A)B^{-1},
$$
from which we can derive:
\beq\label{eme}
  G_{kk'}(\omega+i0) -  G_{kk'}(\omega-i0)
    =  2\pi i \lambda^2 \sum_{l,m=1}^N G_{kl}(\omega+i0) f_l(\omega)
               f_m(\omega) G_{mk'}(\omega-i0) \, .
\eeq
Using the asymptotics:
\beq
  G_{k k'}(\omega)
    \longrightarrow 
    \!\!\!\!\!\!\!\!\!\!\!\!\! _{\stackrel{}{\omega \rightarrow \infty}}
    \frac{\delta_{k k'}}{\omega-\omega'} 
  + o\left(\frac{1}{\omega-\omega'}\right),
\eeq
we prove other useful relations for $G$:
\beq \label{sumo}
  G_{k k'}(\omega \pm i0) 
    = \lambda^2\inid\omega' \sum_{l,m=1}^N 
      f_l(\omega')f_m(\omega') 
      \frac{G_{kl}(\omega'+i0) G_{mk'}(\omega' -i0)}
             {\omega'-\omega \mp i0} \, ,
\eeq
and 
\beq \label{sum}
  \lambda^2 \inid\omega \sum_{l,m=1}^N 
        f_l(\omega)f_m(\omega) 
        G_{kl}(\omega+i0) G_{mk'}(\omega -i0)
   = \delta_{k k'} .
\eeq
Due to the completness of the new basis (\ref{norm})
the old basis vectors may be expressed in terms of the new ones as:
\beq \label {kpsi}
  |k\ra 
  = \inid\omega |\Psi_\omega \ra_{\rm in} \,_{\rm in}\la\Psi_\omega|k\ra  \, , \qquad
  |\omega\ra
    = \inid\omega' |\Psi_{\omega'} \ra_{\rm in} \,_{\rm in}\la\Psi_{\omega'}|\omega\ra
\eeq
where $_{\rm in}\la\Psi_\omega|k\ra$ and $_{\rm in}\la\Psi_{\omega'}|\omega\ra$
are the complex conjugates of $\la k|\Psi_\omega\ra_{\rm in}$ and  
$\la \omega|\Psi_{\omega'}\ra_{\rm in}$ respectively,
which may be obtained from (\ref{evpsol}):
\beqa \label{kpsi1}
  \la k|\Psi_\omega\ra_{\rm in}
  & = & -\lambda \sum_{l=1}^N f_l(\omega) G_{kl}(\omega+i0) \, , \\
  \la \omega|\Psi_{\omega'}\ra_{\rm in}
  & = & \delta(\omega-\omega') - \sum_{k,l=1}^N
        \frac{\lambda^2f_k(\omega)f_l(\omega')G^-_{k,l}(\omega')}
             {\omega'-\omega-i0} \, .
\eeqa
Inserting complex conjugate of (\ref{kpsi1}) into (\ref{kpsi})  we obtain
the inverse relation in the form:
\beqa\label{inverse}
  |k\ra 
  & = & - \lambda \sum_{l=1}^N \int\limits_0^\infty d\omega
          f_l(\omega) G_{kl}(\omega-i0)|\Psi_\omega\ra_{\rm in} \, \\
  |\omega\ra 
  & = & |\Psi_\omega\ra_{\rm in}
      - \sum_{k,l=1}^N \lambda f_k(\omega)\inid\omega' 
        \frac{\lambda f_l(\omega')G^-_{k,l}(\omega')}
             {\omega'-\omega-i0} |\Psi_{\omega'} \ra_{\rm in} \, .
\eeqa
These inverse relations will be used for the calculation
of the time evolution of  $|k\ra$ and $|\omega\ra$ in the next section.

\section {Time evolution}

Using the known evolution of the state $|\Psi_\omega \ra_{\rm in}$, 
$$
e^{-iHt}|\Psi_\omega \ra_{\rm in} = e^{-i\omega t}|\Psi_\omega \ra_{\rm in},
$$
we can find the evolution of the eigenstates of $H_0$:
\beqa \label{timek}
  |k\ra_t
  & = & - \lambda \sum_{l=1}^N \int\limits_0^\infty d\omega\,e^{-i\omega t}
          f_l(\omega) G_{kl}(\omega-i0)|\Psi_\omega \ra_{\rm in} \, ,  \\
\label{timeo}
  |\omega\ra_t
  & = & e^{-i\omega t}|\Psi_\omega\ra_{\rm in}
       - \sum_{k,l=1}^N \lambda f_k(\omega)\int\limits_0^\infty d\omega'
         e^{-i\omega' t}\frac{\lambda f_l(\omega')G^-_{kl}(\omega')}
           {\omega'-\omega-i0} |\Psi_{\omega'} \ra_{\rm in} \, .
\eeqa
Using (\ref{evpsol}), we obtain the representation
\beqa \label{time1k}
  |k\ra_t
  & = & \sum_{l=1}^N A_{kl}(t)|l\ra
      + \lambda \sum_{l=1}^N 
        \inid\omega f_l(\omega) g_{kl}(\omega,t)|\omega\ra  \, , \\
  |\omega\ra_t
  & = & e^{-i\omega t}|\omega\ra  - \lambda^2 \sum_{k,l=1}^N f_l(\omega)
        \inid\omega'f_k(\omega)\frac{g_{kl}(\omega',t)-g_{kl}(\omega,t)}
                                    {\omega'-\omega} \nonumber \\
\label{time1o}
  & + & \sum_{k,l=1}^N \lambda f_k(\omega) g_{kl}(\omega,t)|l\ra 
\eeqa
in terms of the time-dependent matrix functions $A_{kl}(t)$ and $g(\omega,t)$:
\beqa \label {A}
  A_{kl}(t) 
  & = & \lambda^2 \sum_{l,m,n=1}^N \inid\omega e^{-i\omega t} 
  f_m(\omega)f_n(\omega) G_{km}(\omega+i0) G_{ln}(\omega-i0)|l\ra , \\ [2ex]
\label {gsmall}
  g_{kl}(\omega,t)
  & = & - e^{-i\omega t} G_{kl}(\omega-i0) \\ 
  & + & \lambda^2 \sum_{m,n=1}^N \inid\omega'e^{-i\omega' t}
        \frac{f_m(\omega') f_{n}(\omega')G_{km}(\omega'-i0) G_{l n}(\omega'+i0)}
               {\omega'-\omega+i0}  \,  . \nonumber
\eeqa
With the help of (\ref{eme}), we can rewrite (\ref{A}) in the form
\beq \label{A1} 
  A_{kl}(t) 
    = \frac{1}{2\pi i} \inid\omega e^{-i\omega t}
    \left(G_{kl}(\omega+i0)-G_{kl}(\omega-i0)\right)
  = \frac{1}{2\pi i}  \int\limits_C {\rm d}\omega 
    e^{-i\omega t}G_{kl}(\omega) \, ,
\eeq
where the contour $C$ is shown in Fig.~1.
With the help of (\ref{sumo}), we rewrite (\ref{gsmall}) in the form
\beq \label{gsm}
g_{kl}(\omega,t) 
   =  \lambda^2 \sum_{m,n=1}^N \inid\omega' f_m(\omega') f_{n}(\omega')
        G_{km}(\omega'-i0) G_{l n}(\omega'+i0)            
        \frac{e^{-i\omega' t}-e^{-i\omega t}}{\omega'-\omega+i0} \, .
\eeq
The integrand in (\ref{gsm}) does not have any singularity at 
$\omega'=\omega$, therefore $i0$ in the denominator becomes redundant. 
Then using (\ref{eme}) we obtain 
\beqa 
  g_{kl}(\omega,t)
  & = & \frac{1}{2\pi i}\inid\omega'
        \left(G_{kl}(\omega'+i0) - G_{kl}(\omega'-i0)\right)
        \frac{e^{-i\omega' t}-e^{-i\omega t}}{\omega'-\omega} \nonumber \\
\label{gsmall1}
  & = & \frac{1}{2\pi i}\int\limits_C{\rm d}\omega'G_{kl}(\omega')
        \frac{e^{-i\omega' t}-e^{-i\omega t}}{\omega'-\omega} \, ,
\eeqa
where the contour $C$ is shown in Fig.~1. 
For real $\omega>0$ the term with the factor $e^{-i\omega t}$ 
vanishes because it does not have any singularities 
outside the positive part of the real line.
Then we have
\beq \label{gsmall2}
  g_{kl}(\omega,t)
    = \frac{1}{2\pi i}\int\limits_C{\rm d}\omega'G_{kl}(\omega')
      \frac{e^{-i\omega' t}}{\omega'-\omega}  \, .
\eeq
One can easily check the following relation between 
$A_{kl}(t)$ and $g_{kl}(\omega,t)$:
\beq \label{Ag}
  A_{kl}(t)
    = \left(i\frac{\rm d}{{\rm d}t} - \omega\right)g_{kl}(\omega,t) \, .
\eeq
The time evolution of a state $|\Phi\ra$, which is a superposition
of the eigenstates of $H_0$,
\beq \label{ini}
  |\Phi\ra = \sum_{k=1}^N a_k|k\ra ,
\eeq
may be obtained with the help of (\ref{time1k})
\beq \label{init}
  |\Phi(t)\ra = \sum_{k=1}^N a_k|k\ra_t .
\eeq
The survival amplitude $A(t)$ of this state is
\beq \label{samp}
  A(t) 
    \equiv  \la\Phi|\Phi(t)\ra
    = \sum_{k,k'=1}^N a_k a^*_{k'}\la k|k'\ra_t
    = \sum_{k,k'=1}^N a_k a^*_{k'} A_{kk'}(t) \, .
\eeq
Changing the contour of the integration $C$ to $C_1$ in $A_{kk'}(t)$
as shown in Fig.~1, we arrive at
\beq \label{samp2}
  A_{kk'}(t)
    = -\sum_j r^j_{kk'} e^{-iz_j t}
    + \frac{1}{2\pi i } \int_{C_1}{\rm d}\omega e^{-i\omega t} G_{kk'}(\omega),
\eeq
where $r^j_{kk'}$ is the residue of $G_{kk'}(\omega)$ at the pole $z_j$:
\beq \label{rkk}
  r^j_{kk'} = {\rm res} G_{kk'}(\omega)|_{\omega=z_j} \, .
\eeq
The first term in (\ref{samp2}) corresponds to the contribution 
of the poles $z_j$ while the second term is the background integral,
which gives rise to so-called long tail behavior~\cite {Khal,P2}.
It is known that the integral term plays essential role for
very long as well as very short times. In the case of very short times we 
have the well-known Zeno and anti-Zeno regions~\cite {NPN,AKPY,kofman2,FGMR}. 
If we consider the intermediate ``exponential decay'' era, the integral 
term can be neglected because in this time scale,
it is of the next order in $\lambda^2$ compared with the first term.

The same result for $A_{kk'}(t)$ (\ref{samp2}) is obtained in Appendix~A
in terms of Gamov vectors (\ref{surext2}). 
In the intermediate ``exponential'' era, the main contribution to the
survival probability comes from the Gamov vectors as
one may neglect the integral term arising from the background.

\section {Two level model}

The rich structure of the model involving more than one level, will be 
first illustrated with example with two excited levels by choosing the 
formfactor in the form similar to~\cite {LP}
\beq \label{ff}
f_k(\omega)={\omega^{1/4} \over \omega+\rho_k^2} .
\eeq
For this formfactor the matrix element $G^{-1}_{k k'}(\omega)$ (\ref{eta}) is
\beq \label{etaf}
  G^{-1}_{k k'}(\omega)
    = (\omega_k - \omega)\delta_{kk'} 
      + \frac{\pi \lambda^2}{\rho_k+\rho_{k'}}
        \frac{1}{(\sqrt{\omega}+i\rho_k)(\sqrt{\omega}+i\rho_{k'})} \, ,
\eeq
where the first sheet of the complex $\omega$ plane corresponds to the 
upper half of the complex $\sqrt{\omega}$ plane. The square root is defined 
with the cut $[0,+\infty)$ such that $\sqrt{\omega} > 0$ at the upper rim 
of the cut. For $\rho_k >0 $ all singularities of the integral in 
expression~(\ref{eta}) are on the second sheet. 
In the case of two levels the matrix is
\beq \label{eta2}
  G^{-1}(\omega)
    = \left(\begin{array}{l l} \displaystyle
                  (\omega_1 - \omega) + \frac{\pi\lambda^2}{2\rho_1
                  (\sqrt{\omega}+i\rho_1)^2} 
                & \displaystyle \frac{\pi\lambda^2}{(\rho_1+\rho_2)
                  (\sqrt{\omega}+i\rho_1)(\sqrt{\omega}+i\rho_2)} \\ \\
                  \displaystyle \frac{\pi\lambda^2}{(\rho_1+\rho_2)
                  (\sqrt{\omega}+i\rho_1)(\sqrt{\omega}+i\rho_2)}
                & \displaystyle (\omega_2 - \omega) + \frac{\pi\lambda^2}
                  {2\rho_2(\sqrt{\omega}+i\rho_2)^2} 
           \end{array}
  \right)
\eeq
The $2\times 2$ matrix representing the partial resolvent is:
\beq \label{etainv}
  G(\omega) = \det G(\omega)
      \left(\begin{array}{l l} \displaystyle
                  (\omega_2 - \omega) + \frac{\pi\lambda^2}{2\rho_2
                  (\sqrt{\omega}+i\rho_2)^2} 
                & \displaystyle - \frac{\pi\lambda^2}{(\rho_1+\rho_2)
                  (\sqrt{\omega}+i\rho_1)(\sqrt{\omega}+i\rho_2)} \\  \\
                  \displaystyle - \frac{\pi\lambda^2}{(\rho_1+\rho_2)
                  (\sqrt{\omega}+i\rho_1)(\sqrt{\omega}+i\rho_2)}
                & \displaystyle (\omega_1 - \omega) + \frac{\pi\lambda^2}
                  {2\rho_1(\sqrt{\omega}+i\rho_1)^2} 
           \end{array}
  \right) .
\eeq
The determinant $\det G(\omega)$ is
\beqa\label{det}
   (\det G(\omega))^{-1}
   & = & \left[\omega_1-\omega + \frac{\pi\lambda^2}{2\rho_1
               (\sqrt{\omega}+i\rho_1)^2}
         \right] 
         \left[\omega_2-\omega + \frac{\pi\lambda^2}{2\rho_2
               (\sqrt{\omega}+i\rho_2)^2}
         \right] \\
   & - & \left(\frac{\pi\lambda^2}{(\rho_1+\rho_2)
                     (\sqrt{\omega}+i\rho_1)(\sqrt{\omega}+i\rho_2)}
         \right)^2 \, . \nonumber
\eeqa
Here we can formulate necessary conditions for the analyticity of the
function $G^{-1}_{kk'}$ on the first sheet:
\equation\label{conddet2}
   \begin{array}{l l}
      1. & \omega_1\rho_i^2 -\frac{\pi \lambda^2}{2\rho_i} >0 , \quad i=1,2 \\
      2. & \left(\omega_1\rho_1^2-\frac{\pi \lambda^2}{2\rho_1}\right)
           \left(\omega_1\rho_1^2-\frac{\pi \lambda^2}{2\rho_1}\right)
         > \left(\frac{\pi\lambda^2}{\rho_1+\rho_2}\right)^2 \, .
   \end{array}
\endequation
These conditions are definitely satisfied in the weak coupling regime,
because $\omega_1$, $\rho_i$, and $\lambda$ are independent parameters and 
for any fixed $\omega_1$ and $\rho_i$, in the limit $\lambda\rightarrow 0$ 
(\ref{conddet2}) becomes 
$$
\label {conddet3}
   \begin{array}{l l}
      1. & \omega_1\rho_i^2 >0 , \quad i=1,2 \\
      2. & \omega_1\omega_2\rho_1^2\rho_2^2  > 0 \, .
   \end{array}
$$
which is obviously true as $\omega_j$ and $\rho_j$ are positive for any $i$.

In order to find out the analytic structure of $G(\omega)$,
we analyse the poles of the determinant:
\beq \label {detx}
   (\det G(\omega))^{-1}
     = \left[(\omega_1+x^2)(x+\rho_1)^2 - \frac{\pi\lambda^2}{2\rho_1}\right]
       \left[(\omega_2+x^2)(x+\rho_2)^2 - \frac{\pi\lambda^2}{2\rho_2}\right]
     - \left(\frac{\pi\lambda^2}{(\rho_1+\rho_2)}\right)^2 = 0,
\eeq
where we substitute $\sqrt{\omega} = i x$.
This is an algebraic equation of 8$^{\rm th}$ degree with real coefficients, 
so all the roots of this equation are either real or complex conjugated 
pairs. All roots are on the second Riemann sheet, and there can be $k$ 
($k=0\ldots 4$) pairs of complex conjugated roots and $(8-2k)$ real roots 
corresponding to virtual states, i.e. negative energy states on the second 
sheet. For weak coupling $\lambda \rightarrow 0$, the third possibility is 
realized and we have two pairs of complex conjugated roots $z_j$, $z_j^*$, 
which can be evaluated perturbatively as:
\beqa
\label {poles}
 z_j & = & \omega_j + \frac{\pi\lambda^2}{2\rho_j}
     \frac{(\sqrt{\omega_j}-i\rho_j)^2}{(\omega_j+\rho_j^2)^2} + 
     {\pi^2 \lambda^4 \over (\sqrt{\omega_j}+i\rho_j)^2} 
     \left( {1 \over (\omega_j-\omega_k)(\rho_1+\rho_2)^2 
     (\sqrt{\omega_j}+i\rho_k)^2} \right. \nonumber \\
     & - & \left.{1 \over 4\rho_j^2 \sqrt{\omega_j}(\sqrt{\omega_j}+i\rho_j)^3}
     \right) + O(\lambda^6)\, ,  \quad j=1,2, \quad k \neq j.
\eeqa
For the weak coupling regime the expressions for the real and imaginary 
parts of $z_j$ are:
\beqa
\tilde\omega_j & = & \mbox{Re} z_j = \omega_j + \frac{\pi\lambda^2}{2\rho_j}
     \frac{\omega_j-\rho_j^2}{(\omega_j+\rho_j^2)^2} + O(\lambda^4)\, ,  
     \quad j=1,2, \nonumber \\
\gamma_j & = & -\mbox{Im} z_j = \frac{\pi\lambda^2\sqrt{\omega_j}}
                              {(\omega_j+\rho_j^2)^2}  + O(\lambda^4)\, ,  
     \quad j=1,2. \nonumber
\eeqa

Neglecting the integral term in the representation~(\ref{samp2}), 
we can write:
\beqa \label {samp3}
  A(t)
  & \approx & \sum_{k,k'=1,2} a_k a^*_{k'} \sum_{j=1,2}
       e^{-\gamma_j t}e^{-i\tilde\omega_j t}r^j_{kk'} \\
  & = & \sum_{k,k'=1,2} a_k a^*_{k'}
        e^{-i\frac{\tilde\omega_1+\tilde\omega_2}{2}t}
        \left \{\left(r^1_{kk'}e^{-\gamma_1t}
                     +r^2_{kk'}e^{-\gamma_2t}\right) \cos{\nu t}
            + i \left(r^1_{kk'}e^{-\gamma_1t}
                     -r^2_{kk'}e^{-\gamma_2t}\right) \sin{\nu t}
        \right \} \nonumber \\
  & = & \sum_{j=1}^2 |a_j|^2 e^{-i z_j t}  
      - \lambda^2 \sum_{j=1}^2 
        \left( {i\pi |a_j^2| e^{-i z_j t} \over  2\rho_j 
             (\rho_j-i\sqrt{\omega_j})^3 \sqrt{\omega_j} } 
        \right. \nonumber \\
  & + & \left.
             {2\pi \mbox{Re} (a_1a_2^*) e^{-i z_j t} \over (\rho_1+\rho_2)
             (\rho_j-i\sqrt{\omega_j})(\rho_l-i\sqrt{\omega_j})
             (\omega_j-\omega_l) } 
        \right)  + O(\lambda^4), \quad l \neq j, \nonumber
\eeqa
where 
$$
  \nu = \frac{\tilde\omega_1-\tilde\omega_2}{2}.
$$
We would like to notice that both expressions~(\ref{poles}) and (\ref{samp3})
contain the term $1/(\omega_k-\omega_l)$ and, therefore, cannot be directly
used in the case of degenerate levels in the initial Hamiltonian $H_0$.
Also, the case of the continuous spectrum of $H_0$ requires a special 
consideration. 

For the initial conditions $a_1=1$, $a_2=0$, the survival amplitude
(\ref{samp3}) does not have any oscillations. However, such oscillations
appear in the next order $\lambda^4$ in expression (\ref{samp3}).
The survival probability $p(t)$ in the lowest order of $\lambda^2$ can 
be now expressed as:
\beq
p(t) = |A(t)|^2 = ||a_1|^2 e^{-\gamma_1 t} + |a_2|^2 e^{-\gamma_2 t} 
e^{-2 i \nu t} |^2.
\eeq
We illustrate the possible behavior of the survival probability in Fig.~2.
One can see that depending on the initial conditions, the decay can either
mimic the behavior of the usual one level model~\cite{AKPY} or display
considerable oscillations.

\section {$N$-level model}

In the weak coupling regime we can also analyze the $N$-level model with 
an arbitrary formfactor $f_k(\omega)$. 
Using the representation~(\ref{eta}), we find 
\beq
(\det G(\omega))^{-1} = \prod_{k=1}^N (\omega_k-\omega)
-\lambda^2 \sum_{k=1}^N I_{kk}(\omega) \prod_{m\neq k}^N (\omega_m-\omega)
+ O(\lambda^4),
\eeq
where
$$
I_{kl}(\omega) = 
\inid \omega' \frac{f_k(\omega')f_l(\omega')}{\omega' - \omega - i0}.
$$
The zeros of this expression give the position of resonances:
\beq
\label {Npoles}
 z_k = \omega_k - \lambda^2 I_{kk}(\omega_k) + O(\lambda^4)
     = \tilde\omega_k-i\gamma_k \, ,  \quad j=1\ldots N .
\eeq
In the first non-trivial order of the perturbation theory with respect 
to $\lambda^2$ we have:
$$
\tilde\omega_k = \omega_k, \quad \gamma_k = \pi\lambda^2 f^2_k(\omega_k).
$$
The partial resolvent $G$ can also be calculated:
\beq
G_{kk'}(\omega) = 
\left(\omega_k-\omega-\lambda^2 I_{kk'}{\omega}\right)^{-1} \delta_{kk'} 
+ O(\lambda^2).
\eeq
From this representation we obtain the expression for 
the residues~(\ref{rkk}):
\beq
  r^j_{kk'} = -\delta_{kk'}\delta_{kj} + O(\lambda^2).
\eeq  
We derive the survival amplitude~(\ref{samp2}) in the first non-vanishing
term of the perturbation expansion with respect to $\lambda^2$:
\beq \label {NlevA}
  A(t) = \sum_{k=1}^N |a_k|^2 e^{-i \omega_k t} 
  e^{-\pi \lambda^2 f^2_k(\omega_k) t} .
\eeq  

In the case of the $N$-level model, the behavior of the survival probability 
is much more complicated than in two level model. In order to illustrate 
this, we plot in Fig.~3 few examples of the survival probability 
corresponding to different initial conditions for the three level model 
with different parameters.
In this case, the behavior is not necessarily "self-similar" even for the 
very slow decay. 
One can see that our curves reproduce fairly well the experimental results. 
Hence we can suggest here an explanation of the results~\cite {Zewail4} 
which does not refer to the semiclassical description invoked in 
paper~\cite {Zewail4}.
Namely, the initial laser impulse creates in the system $NaI$ a wave packet 
which is a superposition of (many) excited states. Then each excited state
decays independently while the common survival probability~(\ref{NlevA})
exhibits a complicated behavior similar to one of Fig.~3 and Figs.~3, 4 
in paper~\cite {Zewail4}. 

In fact, the interference of many decaying states can drastically change 
the decay patterns. In this case, the decay is equally defined by both 
the parameters of the system (energies, widths) and the distribution of 
the initial wavepacket $a_k$. Therefore, the decay profile may mimic 
different non-exponential functions. Such a behavior is illustrated in 
Fig.~4, where we plot the decay of the $(2N+1)$-level system:
\beq
\label {G-energ}
\gamma_k=\gamma=\mbox{const}, \quad 
\omega_k = \omega_0+{k \over N}\Delta\omega,
\quad k=-N \ldots N.
\eeq
The initial distribution $a_k$ is chosen to be the Gaussian one:
\beq
\label {G-init}
a_k = {\tilde{a}_k \over \sum_k \tilde{a}_k^2}, \quad 
\tilde{a}_k = \exp(-({k \over N})^2), \quad k=-N \ldots N.
\eeq
From Fig.~4 we can see that the initial decay is almost independent of $N$. 
The duration of the initial decay is much shorter than the decay time 
$\tau_{dec}=1/\gamma$. The number of repeated peaks decreases as $N$ 
increases. Already for 10 levels, a rather small number for molecular 
systems, the second peak is very far from the region of the initial decay.
While the values (\ref{G-energ}), (\ref{G-init}) are chosen quite arbitrarily
and can only be used for illustrative purposes, we would like to notice that 
the excitation process in experiments similar to~\cite {Zewail4} is usually
well-defined and well-reproduced. Hence the initial wavepacket may also 
be well-correlated.

\section {Conclusions}

Dissociation processes like the dissociation of NaI,
which is a kind of tunneling/decay process, may be described by
the simple quantum mechanical model of the interaction of 
the $N$-level discrete spectrum with the continuous spectrum. 
Already the model with two levels displays
decaying oscillations of the survival probability in the 
``exponenial'' era, while one-level model exhibit the purely
exponential decay. The amplitude of the oscillation
is determined by the initial state, which is a superposition
of two excited levels. The model with tree levels may illustrate 
qualitatively the experimental curve of the NaI dissociation.
In the $N$-level system, the decay is equally defined by both 
the parameters of the system and the distribution of 
the initial wavepacket.


\acknowledgments 
We would like to thank Prof. Ilya Prigogine for helpful discussions. 
This work was supported by the European Commission 
Project No. IST-1999-11311 (SQID).

\appendix
\section {Time evolution in terms of Gamov vectors}

By analytic continuation to the second
sheet, we obtain the extended distributions $G^d_{kl}(\omega \pm i0)$ 
and $1/[\omega-z_k]_+$ defined as functionals, which
act on a suitable test function $h(\omega)$ as:
\beqa \label {Gd}
  \inid\omega h(\omega)G^d_{kl}(\omega \pm i0)
  & \equiv & \int_\Gamma h(\omega)G_{kl}(\omega \pm i0) \\
  & = & \inid\omega h(\omega)G_{kl}(\omega \pm i0) 
        + 2\pi i \sum_j\int_{C_{z_j}}h(\omega)G_{kl}(\omega \pm i0), \nonumber \\
\label {zop}  
  \inid\omega \frac{h(\omega)}{\left[\omega - z_j\right]_+}
  & \equiv & \int_\Gamma \frac{h(\omega)}{\omega - z_j} 
   = \inid\omega  \frac{h(\omega)}{\omega - z_j}
        + 2\pi i \int_{C_{z_j}} \frac{h(\omega)}{\omega - z_j}.
\eeqa
The contours $\Gamma$ and $C_{z_k}$ are presented in Fig.~1. 
Using (\ref{Gd}) and (\ref{zop}), we obtain from 
(\ref{evpsol}) the Gamov vectors~\cite {PPT,AP,BG} in the form
\beqa \label{Gv}
  |\phi^{\rm G}_j\ra
  & = & N_j\sum_{k,l=1}^N \lambda f_l(z_j)r^j_{kl}
        \left[|k\ra - \inid\omega\frac{\lambda f_k(\omega)}
                          {\left[\omega-z_j\right]_+}|\omega\ra
         \right] \\
\label{tGv}
  \la \tilde\phi^{\rm G}_j|
  & = & N_j\sum_{k,l=1}^N \lambda f_l(z_j)r^j_{kl}
        \left[\la k| - \inid\omega\frac{\lambda f_k(\omega)}
                           {\left[\omega-z_j\right]_+}\la\omega|
         \right] \\
\label{Gvo}
  |\Psi^{\rm G}_\omega\ra 
  & = & |\omega\ra + \lambda\sum_{k,l=1}^N f_l(\omega)G^d_{kl}(\omega + i0) 
        \left\{\int\limits_0^{\infty}{\rm d}\omega'
                \frac{\lambda f_k(\omega')}
                     {\omega'-\omega - i0}|\omega'\ra - |k\ra
               \right\} \\
\label{tGvo}
 \la\tilde\Psi^{\rm G}_\omega|
  & = & \la\omega| + \lambda\sum_{k,l=1}^N f_l(\omega)G_{kl}(\omega - i0) 
        \left\{\int\limits_0^{\infty}{\rm d}\omega'
                \frac{\lambda f_k(\omega')}
                     {\omega'-\omega + i0}\la\omega'| - \la k|
               \right\} \, .
\eeqa
We recall that $r^j_{kl}$ is the residue of $G_{kl}(\omega+i0)$ at the pole 
$z_j$. The normalization constants $N_k$ are:
\beq \label{Nk}
  N_j^{-2} 
    = \sum_{k,l,m,n=1}^N \lambda^2f_l(z_j)f_n(z_j)r^j_{kl}r^j_{mn}
      \left[\delta_{km}+\inid\omega\frac{\lambda^2 f_k(\omega)f_m(\omega)}
                                       {\left[\omega- z_j\right]^2_+}
      \right].
\eeq
The Gamov vectors (\ref{Gv}-\ref{tGvo}) are left and right eigenfunctions
of the extended Hamiltonian, which can be written as
\beq \label{extH}
  H^+  
    = \sum_j z_j |\phi^{\rm G}_j\rl \tilde\phi^{\rm G}_j|
    + \inid\omega\,\omega\,|\Psi^{\rm G}_\omega\rl\tilde\Psi^{\rm G}_\omega|.
\eeq
The Gamov vectors form a biorthonormal set:
\beq \label{gbortn}
  \la\tilde\phi^{\rm G}_j|\phi^{\rm G}_{j'}\ra 
    = \delta_{j j'} \, , \qquad
  \la\tilde\Psi^{\rm G}_\omega|\Psi^{\rm G}_{\omega'}\ra 
    = \delta(\omega-\omega') \, , \qquad
  \la\tilde\Psi^{\rm G}_\omega|\phi^{\rm G}_j\ra 
    = 0 \, ,
\eeq
which is complete. The completeness follows from the 
extension of (\ref{norm}):
\beq \label{norm1}
  I = \sum_{k=1}^N |\phi^{\rm G}_k\rl \tilde\phi^{\rm G}_k|
    + \inid\omega|\Psi^G_\omega\rl\tilde\Psi^{\rm G}_\omega|.
\eeq

The time evolution of the vector $|k\ra$ in the new extended representation is
\beq \label{ktext}
  |k\ra_t
    = \sum_j e^{-iz_kt}|\phi^{\rm G}_j\rl\tilde\phi^{\rm G}_j|k\ra
      + \inid\omega e^{-i\omega t}
        |\Psi^{\rm G}_\omega\rl\tilde\Psi^{\rm G}_\omega|k\ra.
\eeq
Using (\ref{Gv}-\ref{tGvo}), we express the transition amplitude 
\beqa \label{surext}
  \la k|k'\ra_t
  & = & \sum_j e^{-iz_jt} N_j^2 \sum_{l,l'=1}^N 
        \lambda^2 f_l(\omega)f_{l'}(\omega)r^j_{kl}r^j_{k'l'} \\
  & + & \sum_{l,l'=1}^N\inid\omega e^{-i\omega t}
        \lambda^2f_l(\omega)f_{l'}(\omega)
        G^d_{kl}(\omega+i0) G_{k'l'}(\omega-i0). \nonumber
\eeqa
The integral terms of (\ref{surext}) can be rewritten in the form
\beq \label {surext1}
  \frac{1}{2\pi i}\inid\omega e^{-i\omega t}
      \left(G^d_{kk'}(\omega+i0) - G_{kk'}(\omega-i0)\right).
\eeq
Taking into account that $G^d_{kk'}(\omega+i0)$ implies integration
along the contour $\Gamma^*$, which goes to the second Riemann sheet below
all the singularities of $G_{kk'}(\omega+i0)$, we obtain the transition 
amplitude in the form
\beq \label {surext2}
  \la k|k\ra_t
    = \sum_j e^{-iz_jt} N_j^2\sum_{k,l,k',l'=1}^N 
      \lambda^2f_l(z_j)f_l'(z_j)r^j_{kl}r^j_{k'l'}
    + \frac{1}{2\pi i}\int_{C_1}{\rm d}\omega e^{-i\omega t}G_{kk'} \, ,
\eeq
which must coincide with the result obtained using Friedrichs 
solution~(\ref{samp2}). In order to fulfill this requirement the following 
formula must hold:
\beq \label{rr}
  N_j^2\sum_{k,l,k',l'=1}^N \lambda^2f_l(z_j)f_l'(z_j)r^j_{kl}r^j_{k'l'}
    = - r^j_{kk'}.
\eeq

\begin {center}{\bf Figure captions} \end {center}

Fig. 1. The contours of integration ${\rm C}$ and ${\rm C}_1$.

Fig. 2. The survival probability $p(t)$ for the two level model.
The parameters are chosen to be $\gamma_1=\gamma_2=10^{-3}$,
$\omega_1=1.0$, $\omega_1=1.06$. The initial conditions are 
$a_1=0.5$, $a_2=0$ (the solid line), $a_1=0.5$, $a_2=0.2$ 
(the long-dashed line), $a_1=0.5$, $a_2=0.5$ (the short-dotted line).

Fig. 3. The survival probability $p(t)$ for the three level model.
The parameters are chosen to be $\gamma_1=\gamma_2=\gamma_3=10^{-3}$,
$\omega_1=1$, the initial conditions are $a_1=0.3$, $a_2=0.5$, $a_2=0.3$.
The energies are
$\omega_2=1.04$, $\omega_3=1.15$ (the long-dashed line),
$\omega_2=1.06$, $\omega_3=1.15$ (the solid line),
$\omega_2=1.064$, $\omega_3=1.15$ (the short-dashed line).

Fig. 4. The survival probability $p(t)$ for the $N$-level model.
The parameters are chosen to be $\gamma_k=10^{-3}$,
$\omega_k = \omega_0+k\Delta\omega/N$,
$k=-N \ldots N$, where $\omega_0=1$, $\Delta\omega = 0.1$.
The initial conditions are $\tilde{a}_k = \exp(-(k/N)^2)$, $k=-N \ldots N$.
The results for $N=2$ (the solid line), $N=3$ (the short-dashed line),
and $N=5$ (the long-dashed line) are presented.

\end {document}
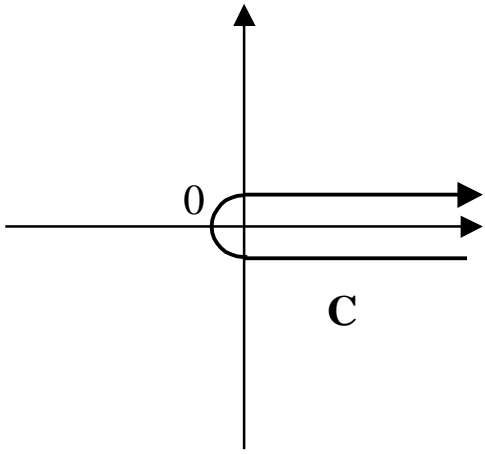
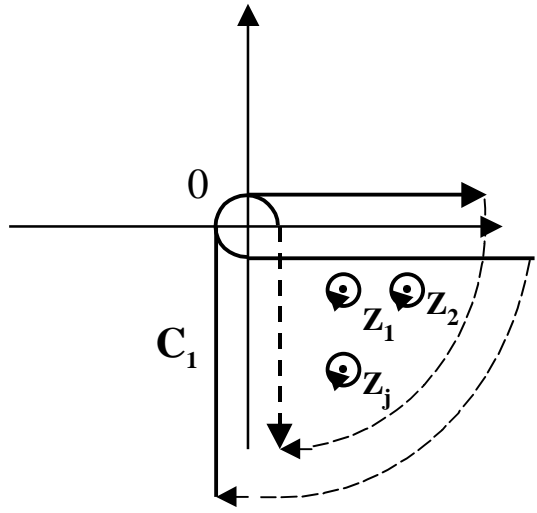
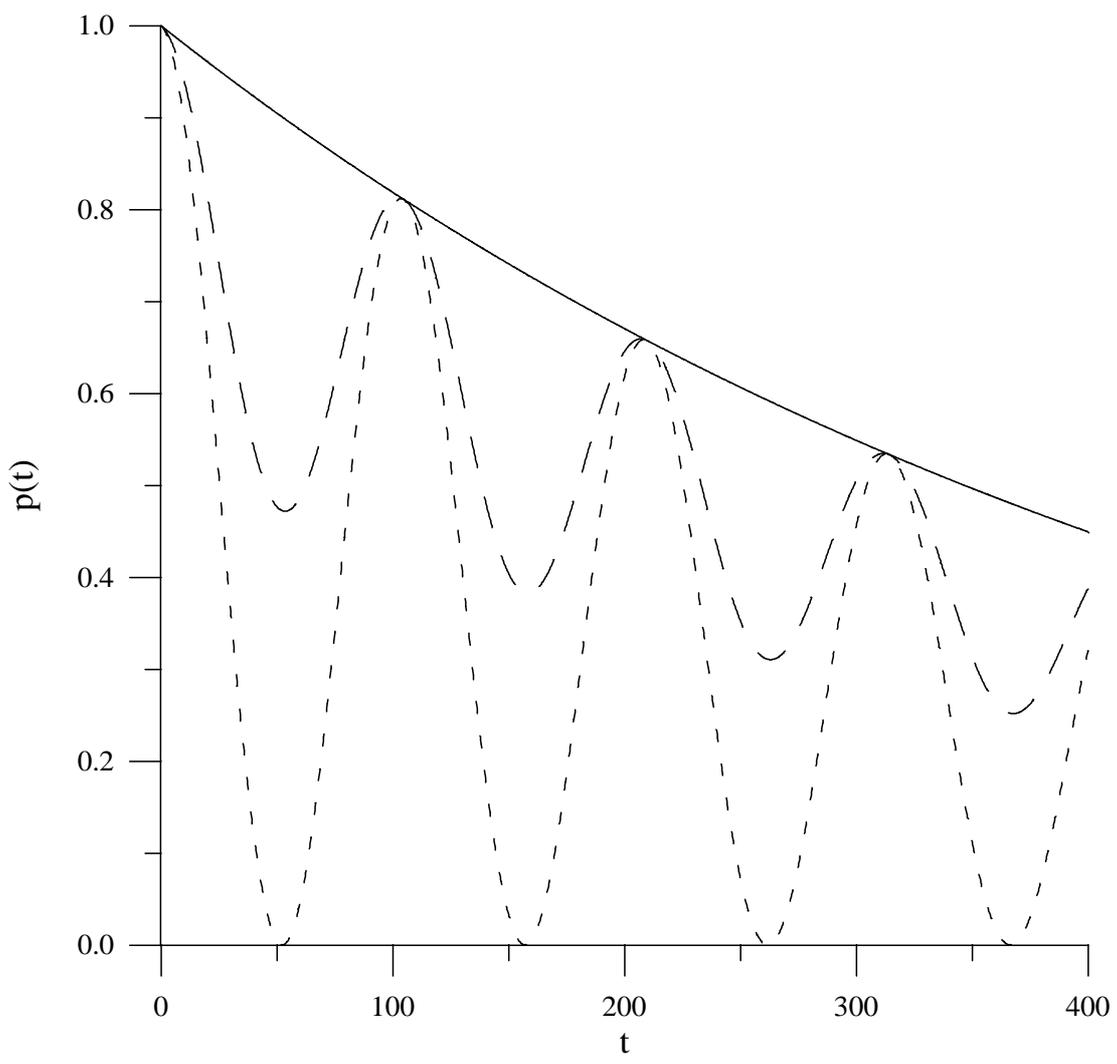
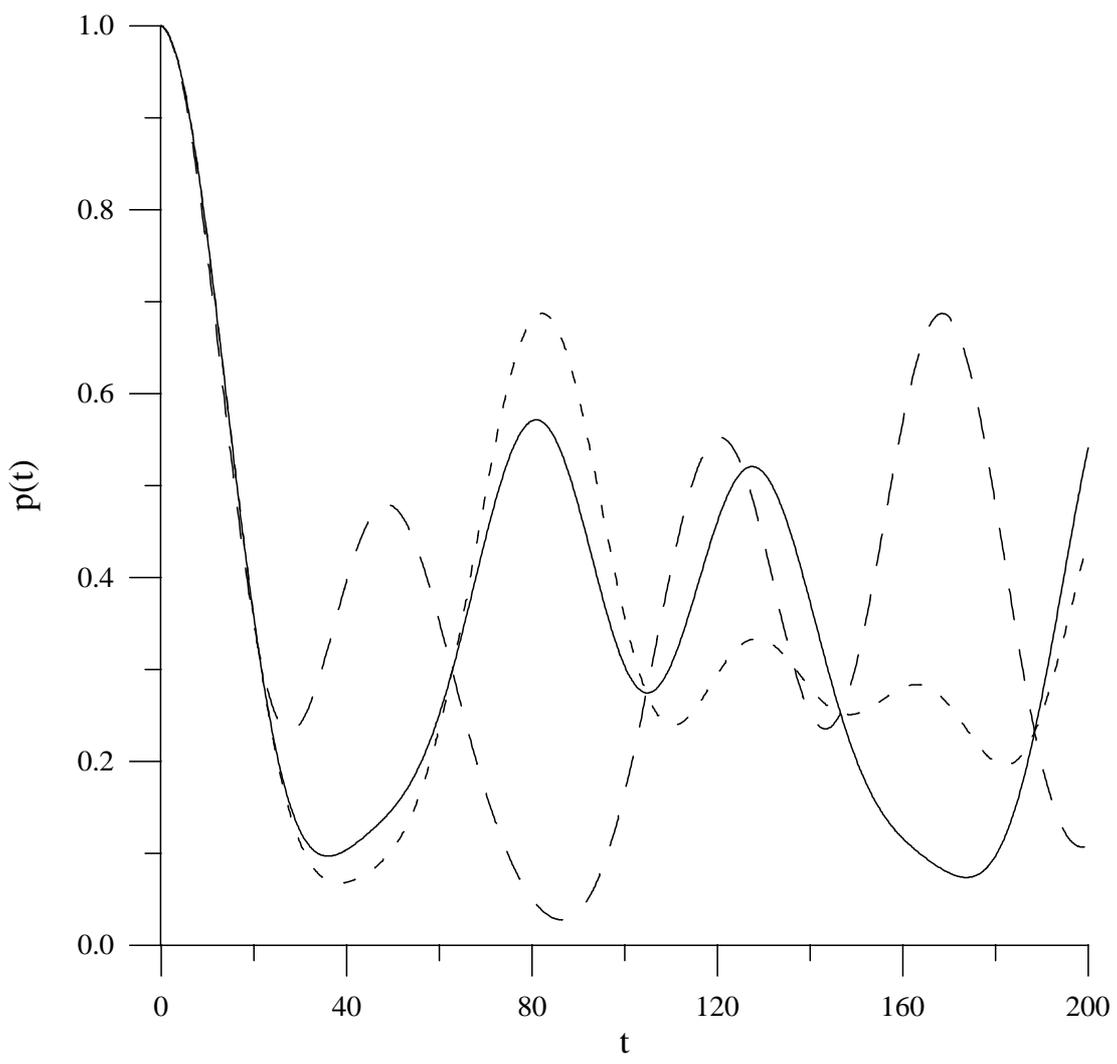
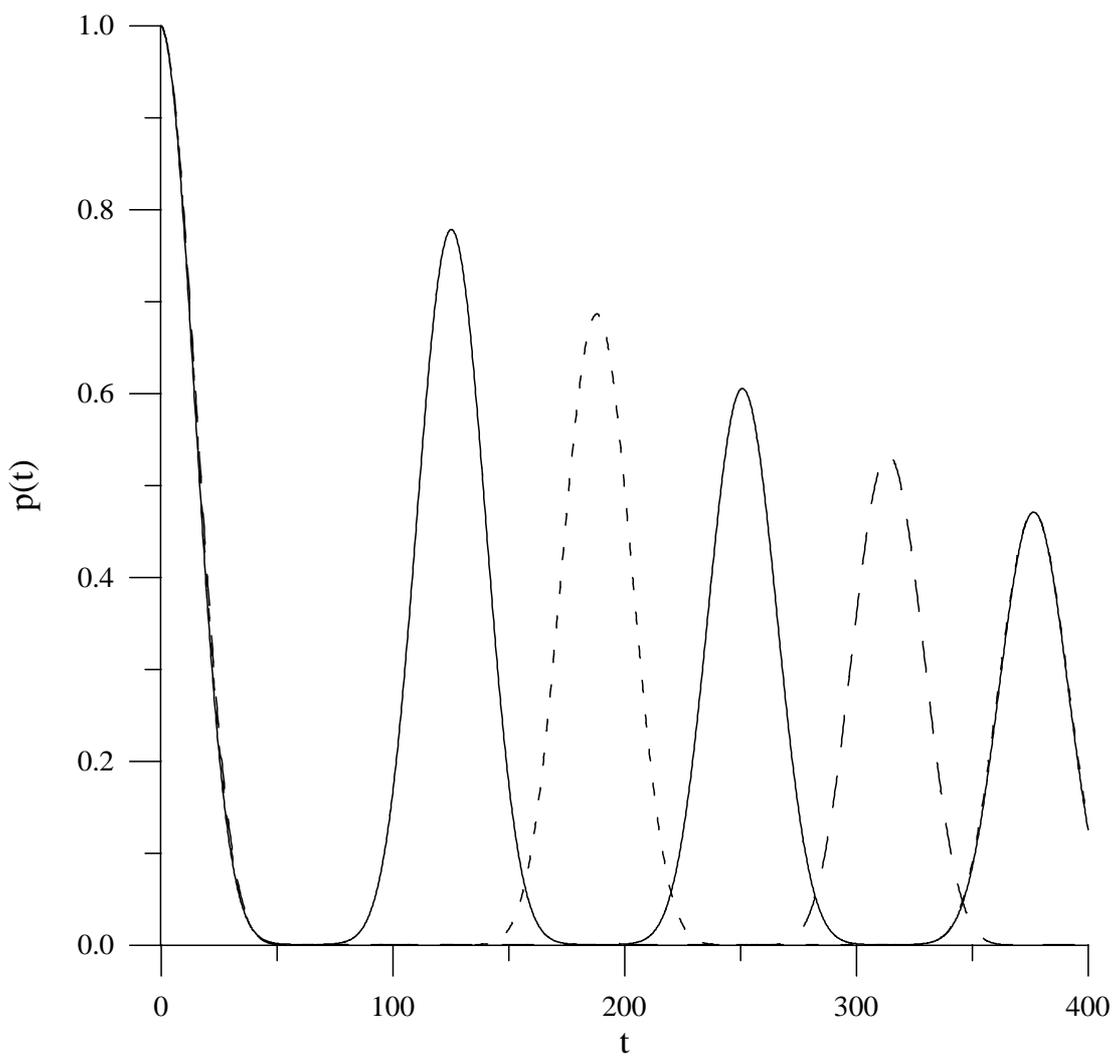
